# Do the solar flares' locations illustrate the boundaries of the solar inner layers?

**Ramy Mawad 1,***

[1] Astronomy & Meteorology Department, Faculty of Science, Al-Azhar University, Nasr City 11488, Cairo, Egypt.

* Correspondence: ramy@azhar.edu.eg

# Abstract

The angular distance of the heliographical or helioprojective coordinates of the solar flares to the projective point of the center of the Sun on the solar disk has been studied during the periods 1975--2021 for GOES events and 2002--2021 for RHESSI events. It gives a specific distribution hereafter Distance. The distribution remains the same without significant changes in the importance value of the solar flare too, with the different coordinate systems, different GOES classifications, and different observational satellites. In addition, it gives the same distribution during each solar cycle. The curvature of the distance distribution has four peaks, which are denoted by the four central rings around the center of the solar disk that look like the solar inner layers in the background. 1) The core circle [0—15°]: it is a projection of the solar core onto the solar disk. 2) Radiative ring [15°—45°]. 3) The convection ring [45—55°]. The limb ring [80°—90°]. A large number of solar flares occurred in the radiative and convection rings. While we have a few events in the core and limb rings.

## 1. Introduction

Although the gaseous nature of the sun allows its interior to be known only through models, other outer layers, such as the photosphere, chromosphere, and corona, can be observed. Most solar phenomena occur in these upper layers, such as solar flares that appear in the chromosphere and photosphere.

Previous studies of the heliographical distribution of solar flares were studied in different methods. Authors of [Gnevyshev, 1967; Shrivastava & Singh, 2005; Zharkova & Zharkov, 2007; Pandey et al., 2015; Abdel-Sattar et al., 2018; Mawad & Abdel-Sattar, 2019] studied the latitudinal distribution. One of the previous studies on the solar flare's location on the solar disk found that the solar flare's latitude varies with the solar activity [Aschwanden, 1994].

Authors of [Jetsu et al., 1996; Cliver et al., 2020, Hongbo et al., 2019; Loumou et al., 2018] studied the longitudinal distribution of solar flares. The latitudinal and longitudinal distributions were studied together, and it was found that the solar flares occur at a specific latitude called the eruptive latitude [Mawad & Abdel-Sattar, 2019]. Most solar flares occur near or within active regions. This is because these solar flares need magnetic energy.

The different layers in the sun can only be seen under certain conditions or in certain bands. except for the innermost layer of the sun's atmosphere, the "photosphere". It is the layer where most of the sun's energy is, and it is always seen. We can observe it directly. Despite being the highest layer, the corona cannot be seen directly. But is it possible to observe the inner layers of the sun, such as the solar core? Is it possible to observe the impact of the inner layers on the solar surface?

The produced energy in the solar core must pass through large amounts of plasma to reach the solar surface, where it is radiated away in mainly two ways: radiation and convection. Solar energy transport switches from radiation to convection. The trans-mission of this energy from the interior to the photosphere and the appearance of interior layers depends on the opacity of the convection zone. Geometrically, we can see the inner layers of the sun. Because the direction of the observer penetrates the inner layers of the sun. Unfortunately, the models give high opacity for the convection zone [Turck-Chièze et al., 1993]. So those inner layers can't be seen. Modern studies such as [Turck-Chièze et al., 1993; Thompson, 2004] have been used in helioseismology [Turck-Chièze & Couvidat, 2011] as an indication of what is inside the sun.

It is known that the position of the heart of the sun, according to the observer, is behind the center of the disk of the sun. Also, the layer of the convective zone is located behind the solar disk in the region near the solar limb. Depending on this, the vertical cross-section of the heliosphere at a heliographical longitude of 90° can simulate the inner solar layers. Therefore, the distance of the solar flare from the center of the solar disk may simulate the inner layers of the sun.

In this study, I will study the angular distance of the solar flares from the center of the solar disk by different coordinate systems, to study whether it simulates the solar inner layers.

## 2. Distance Distribution

As we can see in figure 1 (A) of the horizontal sector of the sun, we have a solar flare point (denoted to $F$) on the spherical surface of the sun. The point of the solar flare has an image in the solar disk in the background (denoted to $F'$), after penetration of the inner layers of the sun. While figure 1(B) shows various locations for solar flares that occur on the solar disk.

The turquoise point is the solar flare occurring above the solar core. The turquoise represents the central distance $D$ of the flare from the center of the solar disk to the flare's location. Consequently, this flare is above the rest of the inner layers. Black and indigo pointer lines represent the central distance of the solar flares above other inner layers such as radiative and convection zone layers. The purple line is the central distance of the solar flare located at the solar limb that equals 90°. It represents the solar photosphere only. It has an angular distance of 1°. While the angular distance is equal to 0 for the exact central flares. Now, we need to distribute the solar flare according to the central angular distance (hereafter, distance or $D$).

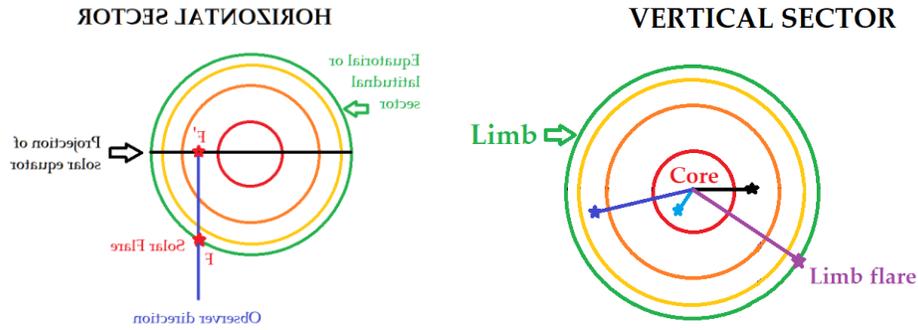

**Figure 1**: Plot (A): Equatorial and latitudinal sectors of the Sun (horizontal sector). The green circle represents the solar latitude. The black circle represents the projection of the solar latitude on the solar disk. *F* is the solar flare on the spherical surface. While *F'* is the projection of the solar flare on the solar disk. Plot (B): The solar disk (vertical sector) of the sun. The turquoise line represents the distance *D* of the flare above the solar core. Consequently, it is above the rest of the inner layers. The black and indigo lines represent the solar flares above other inner layers such as radiative and convection zone layers. The purple line is the central distance of the solar flare above the solar limb. It represents the solar photosphere only.

## 2.1. Distance calculation method

The main vital factor simulating the solar interior layers and projecting them on the solar disk is the distance of the solar flares. The estimation of distance is done by assuming the Sun is a spherical body, using the laws of a spherical triangle, as shown in figure 2, by applying the cosine formula, the distance can be derived by the following formula:

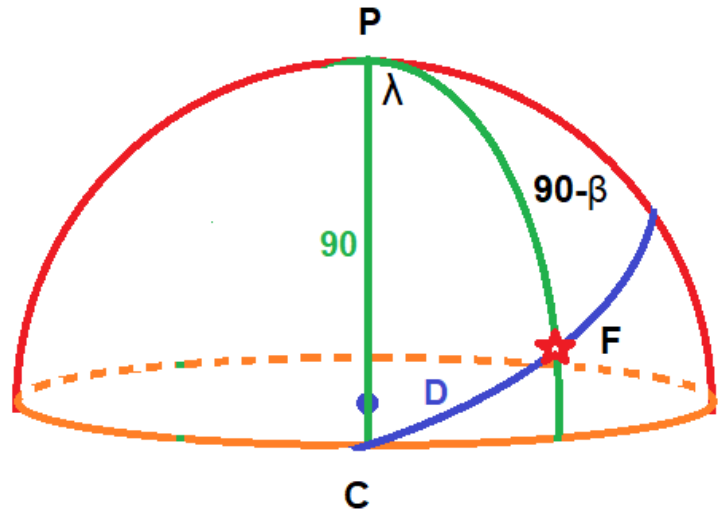

Figure 2: Equatorial and latitudinal sectors of the Sun (horizontal sector). The green circle represents the solar latitude. The black circle represents the projection of the solar latitude on the solar disk. *F* is the solar flare on the spherical surface. While *F'* is the projection of the solar flare on the solar disk. Plot (B): The solar disk (vertical sector) of the sun. The turquoise line represents the distance *D* of the flare above the solar core. Consequently, it is above the rest of the inner layers. The black and indigo lines represent the solar flares above other inner layers such as radiative and convection zone layers. The purple line is the central distance of the solar flare above the solar limb. It represents the solar photosphere only.

$$D = arccos[cos(\lambda) \, cos(\beta)] \qquad (1)$$

Where *D* is the flare's distance between the projected point of the center of the solar disk on the surface of the sphere, and the solar flare position by any coordinate systems such as heliographical or helioprojective coordinates. λ and β are the flare's latitude and longitude, respectively.

We can divide the distance into 90 slices (intervals) to give us a higher accuracy (i.e., 1° interval). This range will be 1—90°. The smallest circle is the closest one to the center which simulates the solar core. The greater one is the circle at the limb.

The RHESSI data is recorded in cartesian coordinates (x, y). We can convert it to the projective coordinates by the following formula, driven by [Mawad et al., 2021].

$$\lambda = arcsin\left[\frac{x}{R_\odot^2 - y^2}\right] \qquad (2)$$

$$\beta = \arcsin\left[\frac{y}{R_\odot}\right] \qquad (3)$$

Where $R_\odot$ is solar radius. Then we can substitute these helioprojective coordinates into equation 1.

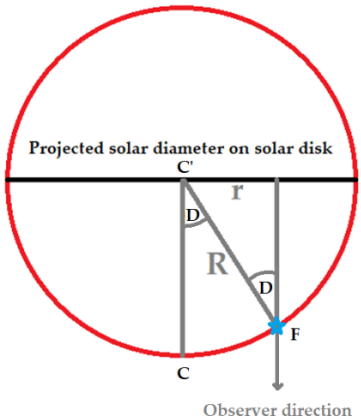

**Figure 3**: The scheme of the great circle $\widehat{CF}$ as shown in figure 2.

## 3. Results and Discussions

The distance was calculated for all solar X—ray (SXR) during the period 1975—2021 for heliographical coordinates obtained by GOES catalog [Mawad et al., 2021], and we counted them according to their distance. In addition to RHESSI solar flares during the period of 2002—2021 for helioprojective coordinates, for comparison purpose.

Figure 3(A) shows the result of the calculated distance for all flares during the selected period with their count of flares at all distances. The behavior of the distance curvature indicates clearly that the flares demonstrate the inner layers.

The central disk events are very few (0 < *D* < 15°). This region reflects the solar core on the inner side. Furthermore, the number of events at the limb (80° < *D* < 90°) is very low, reflecting only photosphere and chromosphere events. while a large number of the flares happened at a distance of about 15°—20°. This region denotes the inner side of the radiative zone after the solar core's projection. Whereas the middle area (20° < *D* < 80°) has a large number of X-ray events, and this is because this region reflects many interior layers in the background.

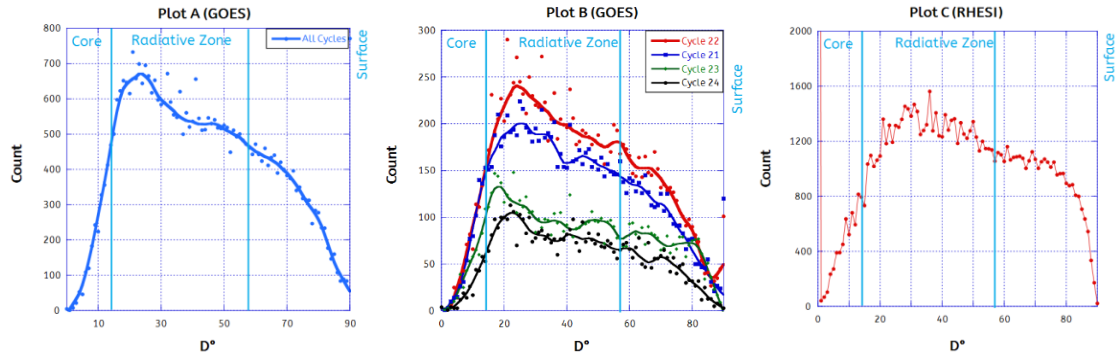

**Figure 4**: The central distance distribution (*D*) of the X-Ray solar flare during all solar cycles (A Panel) and each cycle (B panel) by the flare's count for GOES. While Plot C is for RHESSI.

The data is classified by solar cycle as shown in figure 4(B), to check how solar activity affects the curvature shape. We note that the curvature shape remains the same as it is during each solar cycle (21 to 24 cycles).

Also, we can show about four peaks during the distance range of 0—90°. The main and higher peak (hereafter, the core peak) is a distance of about 15°—20° that reflects the solar core. This means that the small peaks reflect other interior layers, including radiative and convection zones, which we will discuss briefly.

The core's peak moves and changes slightly with time. We notice that the far radius is for cycles 21, 22, 24, and then 23. This means that the solar core radius increases as the strength of the solar cycle progresses and activity increases, and vice versa.

It's worth noting that I repeated the same graph shown in figure 4 but classified data according to flares' GOES classes (B, C, M, and X). Besides, I examined the solar activity by classifying data according to quiet and active periods for all the selected periods and during each cycle. I did not get a significant result. The curvature of figure 4 remains similar.

The plot is applied with RHESSI solar flares during the same solar period as shown in figure 4(C), to check how the solar radiation bands (X-Ray in the current study) and the different coordinate system affect the curvature shape. I note that the curvature shape remains the same. But the results showed a huge number of events (about 25,000 flare events) exactly in the center of the solar disk (*D*=0°) with RHESSI data, unlike those observed by GOES. This is because that the (0, 0) located flares means that these events have not a recorded coordinate in RHESSI catalog. The number of flare events that have a distance greater than 1 is about 90,000.

For additional confirmation of this result, we can calculate the total importance (denoted to I) of these GOES flares that occurred at the same distance.

$$I_D = \sum \frac{f_n}{I} \tag{4}$$

Where $f_n$ is the flare importance value. The solar flare is worth 1 for the X-Class, 10 for the M-Class, 100 for the C-Class, and 1000 for the B-Class. n is the index of flare events that occurred at the same distance *D*. $I_D$ represents the total value of solar flare importance in the X-Class unit that occurred in the distance *D*. Figure 5 shows the compatibility of the total importance with the curvature of the number of events. But the high contrast of the curvature peaks matters more than the count of the events. It is clear that we have 4 peaks similar to figure 4.

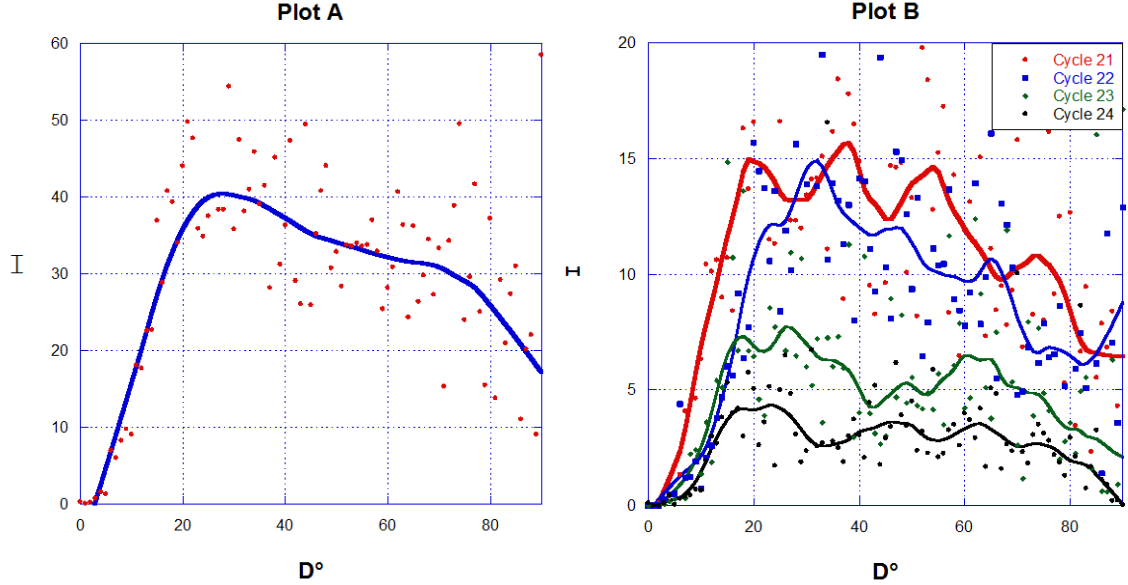

**Figure 5**: The central distance distribution *D°* of the X-Ray solar flare during all solar cycles (A plot) during each cycle (B panel) by the flare's importance, I, in the X-Class unit.

The peaks in the distance curve may be the same as the solar inner layers, or they may represent something else. Therefore, we will distinguish the layers represented on the solar disk by different names inspired by the names of the inner layers, as follows:

- *The core circle*: it is a projection of the solar core on the solar disk.
- *The radiative ring*: it is a projection of the radiative zone on the solar disk.
- *The convection ring*: It is a projection of the convection zone on the solar disk.
- *The limb ring*: It is a projection of the photosphere on the solar disk.

### 3.1. The relationship between distance & radius

Previous studies calculated the radius of the inner layers in units of the radius of the sun $R_\odot$. It differs from the measurement method here used in this study, which reflects the inner layers of the sun. Within the scope of this study, we must convert the radius from scalar distance $R_\odot$ to angular distance *D°*. So that the units are unified, it will be easier to compare the current results with the previous studies. Figure 2 depicts the Sun's great circle $\widehat{CF}$ sector, which is depicted in figure 1. The black line is the projected solar diameter on the solar disk. It may be the solar equator itself if the position of the solar flare is on a solar equator. The distance *D* of any interior layer that has a depth radius of r is given by

$$\sin(D) = r/R \tag{5}$$

By putting the solar radius $R_\odot$ = 1, then

$$D = \arcsin(r) \tag{6}$$

### 3.2. The disk rings and inner layers radius

Figures 4 and 5 show four peaks after the core's peak, including two peaks after the convection zone. Each peak denotes a disk rings, and each has a boundary denoted by two crests. These crests appear clearly in weak solar cycles 23 and 24, especially in solar cycle 23. These peaks overlap during strong solar cycles such as 21 and 22. That longest distance demonstrates the radiative ring. The distance between the core and the radiative rings is not clear because the curve is rising sharply within the solar core.

We already know that X-rays cannot reach easily and directly from the solar core to the surface. However, the increase in the number of solar flares in the region of the core disk, which simulates the core of the Sun, makes us wonder, why? So, I will compare the radius of the solar inner layers to the angular distance of the solar flares. There may prove a correlation or not.

The solar distance of core-radiative zone boundary equals about 0.25 $R_\odot$ according to [Winter & Balasubramaniam, 2014; García et al., 2007; Ryan & Norton, 2010]. According to [Christensen-Dalsgaard, 1991], the radiation-convection boundary occurs at about 0.71 $R_\odot$ Using equation 6, hence,

$$D_c = arcsin(0.25) \simeq 15° \tag{7}$$

$$D_r = arcsin(0.71) \simeq 45° \tag{8}$$

$$D_v = arcsin(0.81) \simeq 55° \tag{9}$$

Where $D_c$, $D_r$, and $D_v$ are the distances from the center to the core, radiative, and convection zones.

This result is consistent with the current results shown in figures 3—5, where the 15° distance represents the core disk (peak of the core).

### 3.3. Distance Model

The first step is to assume the solar surface is a spherical body. The projection of the solar interior layers on the solar disk appears as circles around the center of the solar disk. We can split the Sun into 90 circles centralized by the center of the solar disk which appear as layers around the center of the solar disk. The suggested angular interval between these circles is 1°. We need to calculate the area of these projected circles on the real sphere on the front side and in the background, including the backside too. If we calculated it for the frontside, we could multiply it by n number to give the areas of background spheres, including the backside. The projection of the circles on the real sphere is called "segment", which I want to estimate its area. Each segment has 2 bounders of circles, upper and lower. Each circle has a central angle. θ is for the upper (far) circle and Φ is for the lower (near) circle. Which are measured from any flare's direction (point on this circle) and the center of the solar disk (direction to the Earth in helioprojective coordinates). This solar sphere is depicted in figure 6 in the segment where we want to estimate its area.

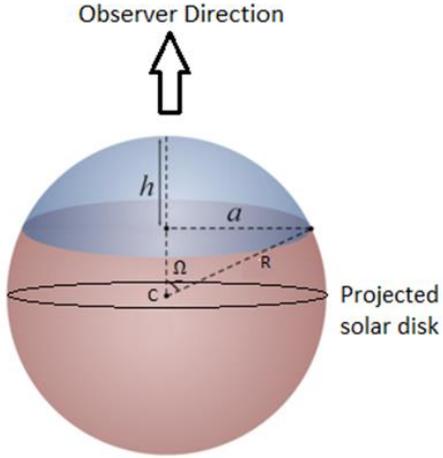

**Figure 6**: The schematic of the Sun. *C* is the center of the Sun and the solar disk. The black circle is the projected solar disk for the Sun for the observer. The spherical cap is the upper boundary of the spherical segment of the solar flares. The difference between the areas of the upper and lower caps gives a segment area.

The area of segment [Donaldson & Siegel, 2001] is the difference between the boundary two caps. We can write,

$$A = 2\pi R_\odot^2 [1 - \cos(\theta)] \tag{10}$$

Where *A* is the area of the frontside segment. $\Omega$ is the angular distance of the projected circle (segment angle). Then, the area of both spherical caps, which have angles θ and (θ +1°) become,

$$A_\theta = 2\pi R_\odot^2 [1 - \cos(\theta)] \tag{11}$$

$$A_{(\theta+1)} = 2\pi R_\odot^2 [1 - \cos(\theta + 1°)] \tag{12}$$

The segment area formula become,

$$A = |\, A_\theta - A_{(\theta+1)}\,| \tag{13}$$

$$A = 2\pi R_\odot^2 [\cos(\theta) - \cos(\theta + 1°)] \tag{14}$$

$$A = 2\pi R_\odot^2 [\cos(\theta) - \cos(\theta)\cos(1°) + \sin(\theta)\sin(1°)] \tag{15}$$

But $R_\odot = 1$, $\sin(1°) \approx \frac{\pi}{180}$, and $\cos(1°) \approx 1$ then,

$$A \approx 2\frac{\pi^2}{180}\sin(\theta) \tag{16}$$

Equation 16 refers to the sinusoidal function. In order to integrate this segment area over all the background layers reaching to the backside of the photosphere, the equation becomes the summation of sinusoids equation [Press et al., 1991] that is written as

$$I_D = v \sum_{n=1}^{m} (a_n \cos(D \times T_n)) \tag{17}$$

Where $a_n$ is the amplitude represented by an inner layer, $T_n$ are the frequencies of angles (period) in degrees. $v$ is the offset value, and D is the distance of the solar flares count *n* or the summation of the importance, I, in X-Class of all solar flares that occurred at the same distance. I set *m* = 3 because this is the best value for the high correlation coefficient and represents the simplest equation.

The compatibility of equation 17 with experimental data was investigated, and it was discovered that equation 16 gives a strong coefficient of determination $R^2$, as shown in table 1. $R^2$ equals 0.97 and 0.752 for the count and the importance of the flares. The sinusoid amplitude was discovered to be greater than the value of $(2\,\pi^2/180)$ in equation 17, indicating that there are many layers in the background added to the front and back sides of the surface. Figures 4 and 5 show the fitting curve.

**Table 1**: The Sum of Sinusoids fitting parameters. $R^2$ is the coefficient of determination. $v$ is the offset value. $a$ is the amplitude. *T* is the period. *P* is the significance of the fit *P-test*. *Chi-sq* is the fitting error.

| N | Count | | Count | |
|---|---|---|---|---|
| | A | T | A | T |
| 1 | -233.9 | 4.408° | -14.84 | 4.408° |
| 2 | -165.8 | 7.804° | -12.13 | 7.804° |
| 3 | -68.5 | 11.053° | -3.895 | 11.053° |
| $v$ | 428.4 | | 28.82 | |
| $R^2$ | 0.97 | | 0.752 | |
| *Chi-sq* | 1.099E05 | | 4059. | |
| *P* | 1.4266E67 | | 2.708E-29 | |

## 4. Conclusions

We study the angular distance of the solar flares from their position in helioprojective and heliographical coordinates to the projection point of the center of the Sun on the solar disk, using the solar flare data obtained from GOES (for heliographical coordinates) during the period 1975—2021, and RHESSI (helioprojective coordinates) during the period 2002—2021.

The solar disk is divided into 90 central rings. The number of solar flares that occurred in each ring was then compiled. It turns out that the results give a definite shape to the oscillation of the distribution.

It has also been noted that the shape of the distance distribution of solar flares remains the same with the different GOES classifications, and with different observations according to the different satellites.

In addition, the different coordinate systems do not give any differentiation in the results. All coordinates give exactly the same distribution.

It was observed that the shape of the curvature did not change when using the overall importance value of the solar flare. But this distribution is now showing the peaks in the curve more than the distribution using the number of solar flares.

There is a fixed number of peaks observed, and each cycle is present. It does, however, sway slightly with each solar activity cycle. It seems to be related to the strength of the solar activity cycle. The number of these vertices is four. Which may mean that the distance distribution may reflect the geometry of the interior of the sun.

It was noted that these peaks form central rings that simulate the inner layers of the sun. So, we divided the solar disk into specific rings in line with the solar inner layers. As these loops simulate and look like the solar inner layers in the background.

The core circle: it is a projection of the solar core onto the solar disk. It has solar flare distances in the range of 0—15°. This ring has very few solar flares for GOES and many for RHESSI. Whereas we do not have solar flares at D=0 except in RHESSI. RHESSI detected more than 25,000 solar flare events at a distance of 0.

The radiative ring: it is a projection of the radiative zone on the solar disk. It has solar flare distances in the range of 15°—45°.

The convection ring: it is a projection of the convection zone on the solar disk. It has solar flare distances in the range of 45°—55°.

The limb ring: it is a projection of the photosphere on the solar disk. It has a very low number of flare events. It has a range of 80°—90°.

A large number of solar flares occurred in the radiative and convection rings. While we have a few events in the core and limb rings.

This result makes us wonder: why does the number of flares differ with the distance from the center? Why are there so few flares on the core disk? Why are there a few flares on the limb disk? Why are there so few flares at the center of the solar disk for GOES but a huge number with RHESSI?

We expect to have a large number at 15° because the distance is a composite of latitude. It is known that most solar flares occur at latitude 15° [Abdel-Sattar et al., 2018; Mawad & Abdel-Sattar, 2019]. But to understand the rest of the distribution requires other studies focused on the distance distribution.

**Acknowledgments**: The author thanks the teams of GOES and RHESSI for supporting the data that helped complete this study.

**Conflicts of Interest**: The author declares no conflict of interest.